\documentclass[a4paper]{jpconf}

\usepackage{amsmath,amssymb,amsfonts,amsthm}
\usepackage{euscript}
\usepackage{graphicx}

\begin{document}

\title{Nonstationary self-gravitating configurations of scalar and electromagnetic fields}

\author{Ju V Tchemarina$^1$, E G Alekseeva$^2$, A N Tsirulev$^{1}$ and N K Nuraliev$^1$}

\address{$^1$ Faculty of Mathematics, Tver State University, 35 Sadovyi, Tver, Russia, 170002}
\address{$^2$ Faculty of Robotics and Complex Automation, Bauman Moscow State Technical University, 5/1 Baumanskaya 2-ya, Moscow, Russia, 105005}

\ead{chemarina.yv@tversu.ru, aeg@bmstu.ru, tsirulev.an@tversu.ru}

\begin{abstract}
Mathematical modeling of gravitating configurations of physical fields is one of the priority directions of the modern theory of gravity. Most of the exact solutions constructed within the framework of the general relativity are static or stationary configurations. This is due to the objective complexity of solving the Einstein equations under the assumption of nonstationarity. We present an approach to constructing nonstationary configurations of a spherically symmetric nonlinear real scalar field and the electromagnetic field, which are assumed both to be minimally coupled to gravity. It is based on the isolation of one invariant equation written in terms of the characteristic function and scalar field potential.
Using the proposed method, an exact nonstationary solution with a nontrivial topology of space-time will be constructed.
\end{abstract}

\section{Introduction}

Nowadays, the construction and study of nonstationary models of astrophysical objects and objects of the microworld based on a self-gravitating scalar field are becoming increasingly popular. One of the current trends in these studies is the use of a scalar field to describe particle-like configurations with nontrivial space-time topology, including spherically symmetric topological geons ~\cite{Dowker1998,BronnikovFabris2006, BronnikovMelnikov2007, Bolokhov2012, Solovyev2012,Kratovitch2017}. It is worth noting that almost all currently known models of topological geons and wormholes are stationary and, accordingly, do not take into account the dynamics of such objects. Also the problem of stability is of great importance when studying configurations with scalar fields.  The stability properties of scalar wormholes under spherical perturbations are the subject of research ~\cite{Bronnikov2018}.

The purpose of this study is to demonstrate a method for constructing nonstationary configurations of a spherically symmetric scalar field with a charge. Using the proposed method, a new exact nonstationary solution will be constructed. This solution is a wormhole whose throat size changes with time.

\section{Action and  basic equations}\label{Sec2}

We start with the action for the gravitating system of nonlinear real scalar field and electromagnetic field which assumed both to be minimally coupled to gravity
\begin{equation}
\Sigma=\int\! \left(-\frac{1}{2}S +\varepsilon\langle d\phi,d\phi\rangle-2V(\phi)-\frac{1}{2}{F}_{{i}{j}}{F}^{{i}{j}}\right) \sqrt[]{|g|}\,d^{\,4}x\,,
\nonumber
\end{equation}
where  $F={F}_{{i}{j}}dx^{i}\wedge dx^{j}$ is the electromagnetic field tensor, $S$ -- the scalar curvature, $\phi$ -- the scalar field, $V(\phi)$ -- the self-interaction potential, $\varepsilon=\pm 1$ -- the sign of the scalar field kinetic term, and the angle brackets denote the scalar product induced by a given space-time metric $g$.

It is clear that the stress-energy tensor is completely specified by the formulas
\begin{equation}
  T = T_{(\phi)} + T_{(em)}\,,
  \nonumber
\end{equation}
\begin{equation}
 T_{{(\phi)}ij}\,=\,2\varepsilon\,\partial_{i}\phi\,\partial_{j}\phi\,-
\,\left(\varepsilon\,g^{km}\,\partial_{k}\phi\,\partial_{m}\phi\,-
\,2\,V\right)\,g_{ij}\,,
\nonumber
\end{equation}
\begin{equation}
T_{{(em)}ij}= -2 g_{ik} F_{jl}F^{kl}+\frac{1}{2} g_{ij}F_{kl}F^{kl}.
\nonumber
\end{equation}

The geometry of space-time is governing by the system of Einstein-Klein-Gordon-Maxwell equations
\begin{equation}\label{En}
\mathcal R_{ij}\,-\frac{1}{2}\,S\,g_{ij}\,=\,T_{ij}\,,
\end{equation}
\begin{equation}\label{KG}
\frac{1}{\sqrt{|g|}}\,\partial_{i}\,(\sqrt{|g|}\,g^{ij}\,\partial_{j}\phi)\,
+\,\varepsilon\,V'_\phi\,=\,0\,,\quad |g|\,=\,|det\,(g_{ij})|\,,
\end{equation}
\begin{equation}\label{max}
\frac{1}{\sqrt{|g|}}\frac{\partial}{\partial x^{i}}(\sqrt{|g|}F^{ij})=0\,.
\end{equation}

The spherical symmetry of space-time allows us to write the electromagnetic field tensor in the following form
\begin{equation}
F=F_{tr}dt\wedge dr,\quad
\text{where}\quad F_{tr}=F_{tr}(t,r)\,.
\nonumber
\end{equation}

Solving equation (\ref{max}) in the coordinate basis $(t,r,\theta,\varphi)$, in which the metric takes the form
\begin{equation}\label{metric1}
g=A^{2}dt\otimes dt-B^{2}dr\otimes dr-C^{2}(d\theta\otimes
d\theta+ \sin^{2}\theta\, d\varphi\otimes d\varphi)\,,
\end{equation}
we obtain
\begin{equation}
F=\frac{A B q}{C^{2}}dt\wedge dr\,.
\nonumber
\end{equation}

To simplify calculations, we will make the transition to the orthonormal basis of vector fields and dual basis of one-forms
\begin{equation}
A\,dt=e^{0},\quad B\,dr=e^{1}, \quad C\,d\theta=e^{3}, \quad  C \sin\theta\,d\varphi=e^{4},\,\quad
F=\frac{q}{C^{2}}e^{0}\wedge e^{1}\,.
\nonumber
\end{equation}

In orthonormal basis, we may write the system of Einstein equations
\begin{equation}\label{SphSymEinstein00new}
-2\frac{C_{(1)(1)}}{C}+2\frac{B_{(0)}C_{(0)}}{BC}-\frac{C_{(1)}^{2}-C_{(0)}^2-1}{C^{2}}\,
 =\,\varepsilon\left(\phi_{(1)}^{2}+\phi_{(0)}^{2}\right)+2\,V +\frac{q^{2}}{C^{4}},
\end{equation}
\begin{equation}\label{SphSymEinstein11new}
-2\frac{C_{(0)(0)}}{C}+2\frac{A_{(1)}C_{(1)}}{AC}+\frac{C_{(1)}^{2}-C_{(0)}^2-1}{C^{2}}\,
 =\,\varepsilon\left(\phi_{(1)}^{2}+\phi_{(0)}^{2}\right)-2\,V -\frac{q^{2}}{C^{4}} ,
\end{equation}
\begin{equation}\label{SphSymEinstein10new}
\frac{A_{(1)(1)}}{A}-\frac{B_{(0)(0)}}{B}+\frac{C_{(1)(1)}}{C}-\frac{C_{(0)(0)}}{C}+\frac{A_{(1)}C_{(1)}}{AC}-\frac{B_{(0)}C_{(0)}}{BC}=\,\varepsilon\left(\phi _{(0)}^{2}-\phi_{(1)}^{2}\right)-2\,V +\frac{q^{2}}{C^{4}},
\end{equation}
\begin{equation}\label{SphSymEinstein01new}
-2\frac{C_{(0)(1)}}{C} +
2\frac{B_{(0)}C_{(1)}}{BC}\,\equiv\,-2\frac{C_{(1)(0)}}{C} +
2\frac{A_{(1)}C_{(0)}}{AC}
 =\,2\,\varepsilon\phi _{(0)}\phi _{(1)}\,,\,\,\,
\end{equation}
and Klein-Gordon equation
\begin{equation}\label{ScalarFieldnew}
\phi_{(0)(0)}-\phi_{(1)(1)}+
\phi_{(0)}\frac{(BC^2)_{(0)}}{BC^2}-\phi_{(1)}\frac{(AC^2)_{(1)}}{AC^2}+
\,\varepsilon V'_\phi\,=\,0.
\nonumber
\end{equation}
Lower indexes in round parentheses denote the directional derivatives along the corresponding vector field.

\section{Characteristic function}\label{Sec3}

Following ~\cite{Tchemarina2009,Solovyev2012,Nikonov2016,Kratovitch2017,Kratovich2017}, in an invariant manner, we will define the function
\begin{equation}\label{f}
f=f(\phi,C)\,=\,-\,\left\langle
dC,dC\right\rangle\,=\,C_{(1)}^{2}\,-\,C_{(0)}^{2}\,,\,\,dC=\,C_{(0)}{e^0}\,-\,C_{(1)}{e^1}\,.
\end{equation}
The solutions of equation $f(\phi,C) = 0$ define hypersurfaces on which the 1-form $dC$ becomes  null. For stationary configurations it is true on event horizons and hence the function $f(C,\phi)$ will be referred to as the characteristic function. The asymptotic behavior of this function also makes it possible to assign a solution to a certain type.

We call the scalar field configuration as  nonstationary or stationary according as $\phi\neq\phi(C)$ or $\phi=\phi(C)$. The latter case is well studied. Therefore, we will focus on the consideration of non-stationary configurations. Note that the radius $C$ is defined in an invariant manner by the metric (\ref{metric1}).

Making use of the formula (\ref{f}), we find the derivatives of $f(\phi,C)$ in the directions of basis vector fields
\begin{equation}\label{G22}
f_{(1)}=2C_{(1)}C_{(1)(1)}-2C_{(0)}C_{(0)(1)}\,,\,\,
f_{(0)}=2C_{(1)}C_{(1)(0)}-2C_{(0)}C_{(0)(0)}\,.
\nonumber
\end{equation}
We thus obtain the relation
\begin{equation}
C_{(1)(1)}=\frac{f_{(1)}}{2C_{(1)}}+\frac{C_{(0)}C_{(0)(1)}}{C_{(1)}}\,.
\nonumber
\end{equation}
Substituting the last expression in equation (\ref{SphSymEinstein00new}), we obtain
\begin{equation}
-\frac{2}{C}\left(\frac{f_{(1)}}{2C_{(1)}}+\frac{C_{(0)}C_{(0)(1)}}{C_{(1)}}\right)+\frac{2B_{(0)}C_{(0)}}{BC}-\frac{f-1}{C^{2}}=
\varepsilon\left(\phi^{2}_{(1)}+\phi^{2}_{(0)}\right)+2V+\frac{q^{2}}{C^{4}}\,.
\nonumber
\end{equation}
After multiplication by $C_{(1)}\,C^{2}$, it gives
\begin{equation}\label{G1}
[C(f-1)]_{(1)}=2\frac{B_{(0)}C_{(0)}CC_{(1)}}{B}-2CC_{(0)}C_{(0)(1)}-C_{(1)}\left(C^{2}\varepsilon\left(\phi^{2}_{(1)}+\phi^{2}_{(0)}\right)+2C^{2}V+\frac{q^{2}}{C^{2}}\right)\,.
\end{equation}
From equation (\ref{SphSymEinstein01new}), we find
\begin{equation}\label{G01}
C_{(0)(1)}=\frac{B_{(0)}C_{(1)}}{B}-\varepsilon C\phi_{(0)}\phi_{(1)}\,.
\end{equation}
By combining equations (\ref{G1}) and (\ref{G01}), we obtain equation
 \begin{equation}
[C(f-1)]_{(1)}=C_{(1)}[\varepsilon C^{2}(\phi^{2}_{(1)}-\phi^{2}_{(0)})-2C^{2}V-\frac{q^{2}}{C^{2}}]
+\phi_{(1)}[2C^{2}\varepsilon(\phi_{(0)}C_{(0)}-C_{(1)}\phi_{(1)})]\,.
\nonumber
\end{equation}
By a similar sequence of conversion, we obtain from equation (\ref{SphSymEinstein11new}) the appropriate equation
\begin{equation}
[C(f-1)]_{(0)}=C_{(0)}[\varepsilon C^{2}(\phi^{2}_{(1)}-\phi^{2}_{(0)})-2C^{2}V-\frac{q^{2}}{C^{2}}]
+\phi_{(0)}[2C^{2}\varepsilon(\phi_{(0)}C_{(0)}-C_{(1)}\phi_{(1)})]\,.
\nonumber
\end{equation}
For a non-stationary configuration the scalar field $\phi$ and radius $C$ can be considered as independent variables. Therefore, the last two equations are equivalent to one invariant equation
\begin{equation}\label{Gl_ur}
d[C(f-1)]=C^{2}(\varepsilon\left(\phi_{(1)}^2-\phi_{(0)}^2\right)-2V-\frac{q^2}{C^4})dC+2C^2\varepsilon(\phi_{(0)}C_{(0)}-\phi_{(1)}C_{(1)})d\phi\,,
\end{equation}
where
\begin{equation}
\phi_{(0)}C_{(0)}-\phi_{(1)}C_{(1)}= <d\phi,dC>\,,\quad \phi_{(1)}^2-\phi_{(0)}^2= -<d\phi,d\phi>\,.
\nonumber
\end{equation}
According to equation (\ref{Gl_ur}),
\begin{equation}\label{ur21}
<d\phi,dC>=\frac{\varepsilon}{2C}\,f'_{\phi}\,,\quad <d\phi,d\phi>= -\varepsilon\left(2V+\frac{f'_{C}}{C}\,+\frac{f-1}{C^2}+\frac{q^2}{C^4}\right)\,.
\end{equation}
Thus, we isolate one invariant equation written in terms of the characteristic function and scalar field potential.

\section{Approach to constructing nonstationary configurations}\label{Sec4}

A significant part of the proposed method is the use of coordinates $( \phi, C, \theta, \varphi)$ which is possible, at least locally, for any nonstationary configuration.
Note that the expressions (\ref{f}) and (\ref{ur21}) define metric coefficients
\begin{equation}
g^{\phi \phi}=<d\phi,d\phi>,\quad g^{\phi C}=<dC,d\phi>,\quad\,g^{C C}=<dC,dC>\,.
\nonumber
\end{equation}
Thus, finding the inverse matrix, we obtain the components of the covariant metric tensor and write the metric in coordinates $( \phi, C, \theta, \varphi)$
\begin{equation}
d s^{2}=-4\frac{\varepsilon C^{4} f d\phi^{2}+C^3 f'_{\phi}d\phi dC  +(2C^{4}V + C^3 f'_{C} +C^2( f - 1)+q^2)d C^{2}}{4fC^2(2C^{2}V + C f'_{C} + f - 1)-\varepsilon(Cf'_{\phi})^2+ 4fq^2}-
\nonumber
\end{equation}
\begin{equation}\label{metric20}
-C^{2}\left(d\theta^2+ \sin^{2}\theta\,d\varphi^2\right)\,.
\end{equation}

Next we find the  electromagnetic field tensor in the coordinates $(\phi, C, \theta, \varphi)$ in the form
\begin{equation}
 F=\frac{2q\,d\phi \wedge dC} {\sqrt{-4\varepsilon f C^{2}(Cf'_C+ f-1+ 2C^{2}V)-4\varepsilon q^{2}f +C^{2}{f'_{\phi}}^{2}}}\,.
 \nonumber
\end{equation}

Then we sequentially find the components of the stress-energy tensor and write down the system of Einstein equations (\ref{En}). We do not give the corresponding expressions here because of their bulkiness. It turns out that the resulting system of equations is equivalent to the single equation (\ref{KG})
\begin{equation}
8V^2C^8+(6C^7f'_{C}+8C^6f-2\varepsilon C^6f''_{\phi \phi}+8C^4q^2-8C^6)V+\varepsilon C^6V'_{\phi}f'_{\phi}-f''_{C C}C^6f+
\nonumber
\end{equation}
\begin{equation}
+\varepsilon C^5f''_{\phi C}f'_{\phi}+\varepsilon(C^4-fC^4-C^2q^2-f'_{C}C^5)f''_{\phi \phi}+(f'_{C})^2 C^6+(3C^3q^2+3C^5f-3 C^5)f'_{C}+
\nonumber
\end{equation}
\begin{equation}\label{S10}
+4f^2C^4+(8C^2q^2-6C^4)f+2C^4+2q^4-4C^2q^2=0.
\end{equation}

Thus, the characteristic function and the scalar field potential are connected by one single Klein-Gordon equation. The solution can be completed by integration of
equation (\ref{S10}). However, this equation is too complicated to solve it explicitly, without imposing any additional restrictions. It seems most convenient to find the unknown function ($f(\phi,C)$ or $V(\phi)$) as a series or numerical solution.

\section{Special case for metric function}\label{Sec5}

Despite the complexity, equation (\ref{S10}) in some cases allows to obtain exact analytical solutions.
Choosing the characteristic function in special form, we obtain an explicit solution for the scalar field potential.
Now suppose that
\begin{equation}\label{f1}
f\left( \phi,C \right)=1+C^2 h(\phi),\quad  \varepsilon=-1\,,\quad q=0.
\end{equation}

This choice of the characteristic function is not accidental. It is due to the possibility of obtaining exact solutions with a nontrivial topology of space-time.
The use of a phantom scalar field with a negative kinetic term is necessary for constructing solutions with nontrivial topology.
The impossibility of the existence of topological geons and wormholes without a horizon with a positive kinetic term was proved in ~\cite{Morris1988, Hochberg1997, BronnikovShikin2002}.

The substitution (\ref{f1}) reduces equation (\ref{S10}) to the form
\begin{equation}\label{S11}
8\, V^{2}+ \left( 2\,h''_{\phi \phi}+20 h\right)V-V'_{\phi}\,h'_{\phi}-2\,{h'_{\phi}}^{2}+2 \,h''_{\phi \phi}\,h+12\,{h}^{2}\,=0\,.
\end{equation}

The formal solution of equation (\ref{S11}) can be readily written down in terms of integrals
\begin{equation}\label{S12}
V(\phi)=-\frac{3\,h(\phi)}{2}-\frac{{h'_{\phi}}^{2}}{8\,h(\phi)}- \frac{e^\emph{F}\,{h'_{\phi}}^{2}}{8\,h^2(\phi)\,\int{\frac{e^\emph{F}\,h'_{\phi}}{h^2(\phi)}d\,\phi}}\,,\,\,
\emph{F}(\phi)=-4\int{\frac{h(\phi)}{h'_{\phi}}d\,\phi}\,.
\end{equation}
These integral formulas allow us to recover the scalar field potential by a given characteristic function.

\section{An exact nonstationary solution with a nontrivial topology of space-time}\label{Sec6}

Using the proposed method, we construct a model of nonstationary wormhole. Substituting in the formulas (\ref{S12}) the function $h(\phi)=\,\phi^2\,-\,1$\,,
we find,
\begin{equation}
V(\phi)=\,1\,-\frac{3\,\phi^2}{2}\,+\frac{1\,+\,\phi^2\,-\,e^{\,\phi^2}}{2\left(1\,+\,e^{\,\phi^2}\left(\phi^2\,-\,1\right)\right)}\,.
\nonumber
\end{equation}
The integration constant was chosen so that the scalar field potential on the wormhole throat $\left(\phi=0\right)$ is a regular function.

The metric (\ref{metric20}) now takes the form
\begin{equation}
d s^{2}=\frac{1}{\Delta}\left(\left(C^{2}\left(1\,-\,\phi^2\right)-\,1\right) d\phi^{2}\,+\,2\,\phi\,C\,d\phi\, dC\,+\,\frac{\phi^2\,\left(1\,-\,e^{\,\phi^2}\right)}{1\,+\,e^{\,\phi^2}\left(\phi^2\,-\,1\right)} d C^{2}\right)-
\nonumber
\end{equation}
\begin{equation}
-C^{2}\left(d\theta^2+ \sin^{2}\theta\,d\varphi^2\right)\,,
\nonumber
\end{equation}
where
\begin{equation}
\Delta=\frac{\phi^2\,\left(e^{\,\phi^2}\,-\,\phi^2\,C^2\,-\,1\right)}{1\,+\,e^{\,\phi^2}\left(\phi^2\,-\,1\right)}\,.
\nonumber
\end{equation}

Space-time, in the general relativity, is considered as a manifold with a Lorentzian signature.
Therefore, the solution is defined in the area which is determined by the inequality
\begin{equation}
e^{\,\phi^2}\,-\,\phi^2\,C^2\,-\,1\,<\,0\,.
\nonumber
\end{equation}

Next, we revert to the usual coordinates $( t, r, \theta, \varphi)$.

By letting
\begin{equation}\label{phiC}
\phi\,=\,r\,,\quad C\,=\,\cosh(t)\sqrt{\frac{e^{\,r^2}\,-\,1}{r^2}}\,=\,\cosh(t)\left(1\,+\,\frac{r^2}{4}\,+\,O(r^4)\right)\,,\quad r\rightarrow 0\,,
\nonumber
\end{equation}
the metric reduces to the standard form (\ref{metric1}).

Therefore,
\begin{equation}
A^2=\frac{e^{\,r^2}\,-\,1}{r^2}=1+\frac{r^2}{2}+O(r^4)\,,\quad
B^2=\frac{e^{\,r^2}\left(r^2\,-\,1\right)\,+\,1}{r^2\left(e^{\,r^2}\,-\,1\right)}=\frac{1}{2}+\frac{r^2}{12}+O(r^4)\,,\quad r\rightarrow 0\,.
\nonumber
\end{equation}

\begin{figure}[ht!]
  \centering
  \begin{minipage}{0.4\textwidth}
    \includegraphics[width=\textwidth]{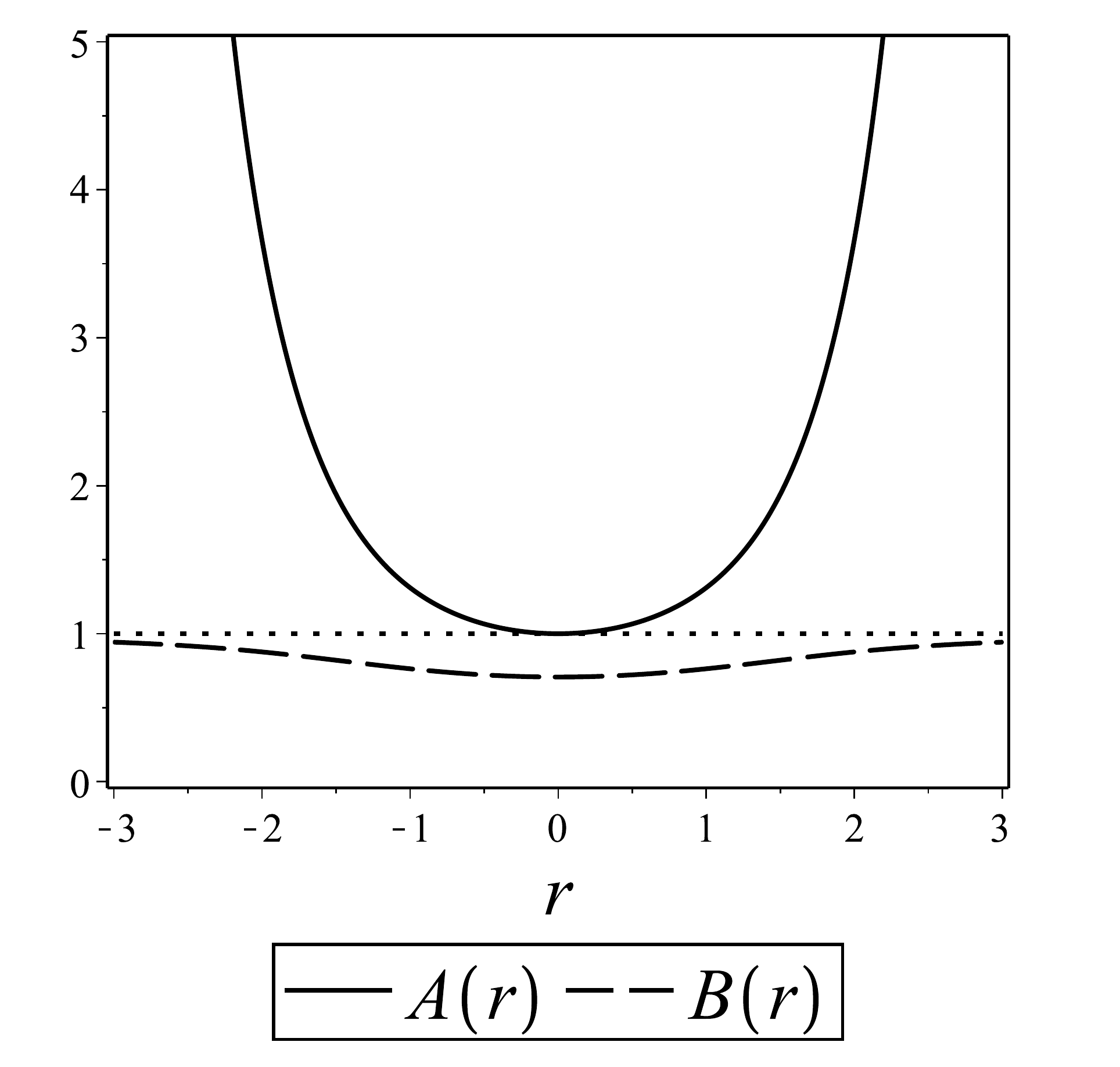}\\
  \end{minipage}\quad\quad\quad
\begin{minipage}{0.4\textwidth}
    \includegraphics[width=\textwidth]{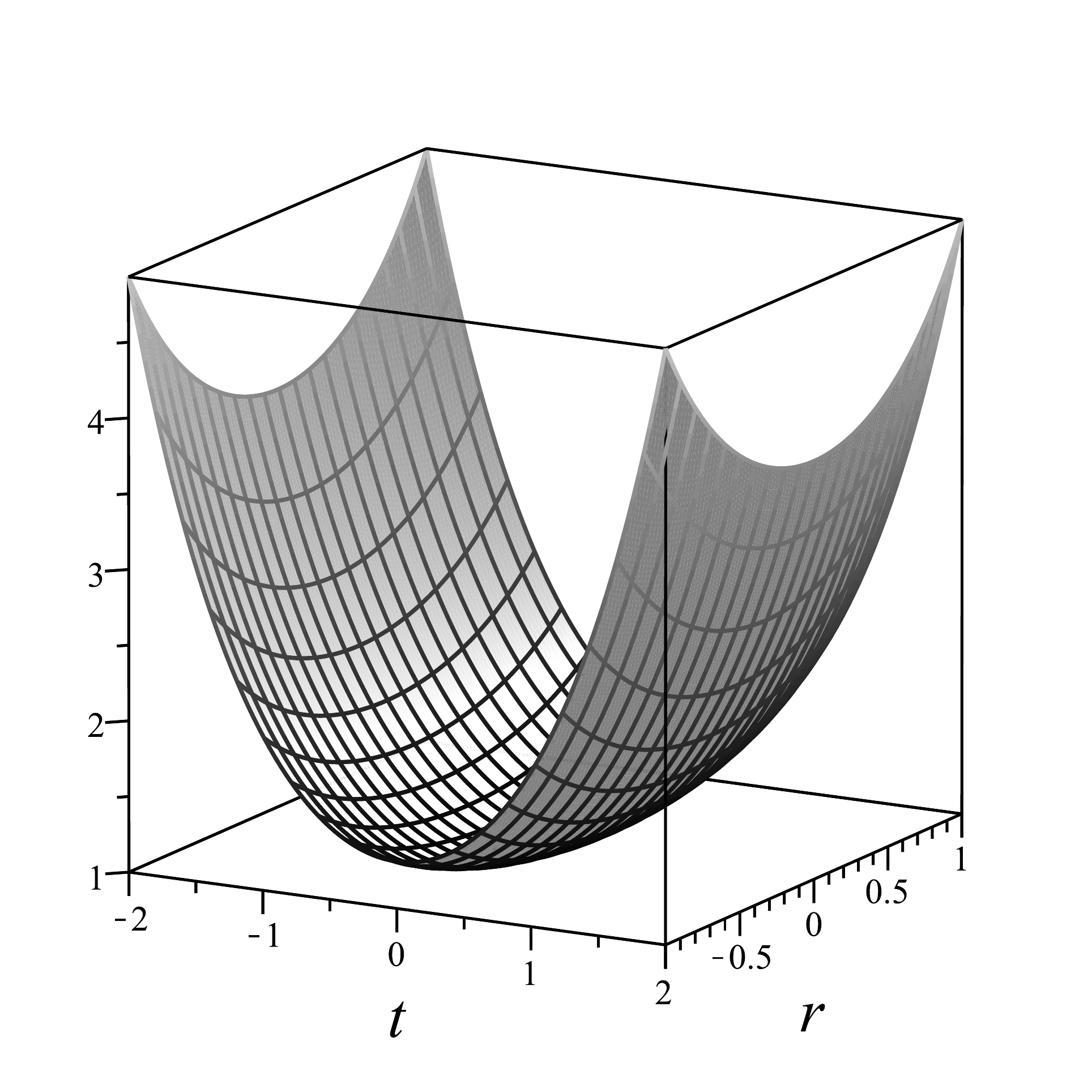}\\
  \end{minipage}
   \caption{Left plot: metric functions $A(r)$ and $B(r)$. Right plot: surface $C(t,r)$.}
  \label{F1}
\end{figure}

The plots of the metric functions $A(r),\,B(r)$ and the surface $C(t,r)$ are presented in Figure~\ref{F1}.
Metric functions are regular, positive and even with respect to $r\,$. The coordinate $r$ is space-like everywhere. This allows the solution to be interpreted as a wormhole, the size of which $C(t,0)\,=\,\cosh(t)$ varies over time, taking the smallest value $C\,=\,1$ at $t\,=\,0\,$.
The constructed model of a spherically symmetric wormhole illustrates the use of new mathematical methods for studying and solving the Einstein's equations.

\section{Conclusions}

We present an approach to constructing nonstationary configurations of a spherically symmetric nonlinear real self-gravitating scalar field and the electromagnetic field. It is based on the isolation of one invariant equation written in terms of the characteristic function and scalar field potential. Despite the fact that the question of a general solution of this equation is open, this approach allows us to test known numerical nonstationary solutions. Under additional conditions, it makes it possible to construct nonstationary solutions with specified properties.

\end{document}